\documentclass[seceq]{ptptex}

\usepackage{graphicx}
\usepackage{wrapft}

\title{%        %You can use \\ for explicit line-break.
Magnetic aspects of QCD and compact stars%
}

\author{%       %Use \scshape for the family name.
Toshitaka \textsc{Tatsumi}%
}

\inst{%     %Affiliation, neglected when [addenda] or [errata].
Department of Physics, Kyoto University, Kyoto 606-8502
}

\abst{%         %This abstract is neglected when [addenda] or [errata].
Magnetic properties of quark matter are discussed. The possibility of ferromagnetic transition is studied by using the one-gluon-exchange interaction. Magnetic susceptibility is evaluated within Landau Fermi liquid theory, and the important roles of the screening for the gluon propagation are elucidated. Static screening for the longitudinal gluons improves the infrared singularities, while the transverse gluons receive only dynamic screening. The latter property gives rise to a novel non-Fermi-liquid behaviour in the magnetic susceptibility.  The critical density is estimated to be around the nuclear density and the Curie temperature several tens MeV.  The spin density wave is also discussed at moderate densities, where chiral transition becomes important. Pseudoscalar condensate as well as scalar one takes a spatially non-uniform form in a chirally invariant way. Accordingly magnetization oscillates like spin density wave. 
These results should have some implications on compact star phenomena.
}

\begin{document}

\maketitle

\section{Introduction}
Nowadays there are many studies about the phase diagram of QCD on temperature-density plane \cite{wam}. Here we'd like to explore some magnetic phases by considering the spin degrees of freedom.  Phenomenologically such magnetisms should be related to observation of compact stars, especially their magnetic evolution. The origin of the strong magnetic field in compact stars is a log-standing problem since the first discovery of pulsars \cite{mag}. Recent discovery of magnetars with huge magnetic field ($10^{15}$G) seems to revive this issue. Their origin is not clear yet, while some ideas such as fossil field, dynamo scenario has been proposed. If QCD has a potential to produce such magnetic filed, it gives a microscopic origin. Actually we studied a possibility of spontaneous magnetization of quark matter \cite{tat00}, since microscopic nuclear-matter calculations have shown negative results \cite{nuc}. We have shown that quark matter would be ferromagnetic state due to the Bloch mechanism, in analogy with electron gas, and suggested that magnitude of the magnetic field amounts to be $O(10^{15-17}G)$ if such state develops inside compact stars. 

We shall discuss a possibility of spin density wave (SDW) phase as another interesting magnetic aspect of QCD in relation to chiral transition \cite{dcdw}. It is well known that restoration of chiral symmetry is important at moderate densities and many efforts have been devoted to figure out their properties. Assuming the non-uniform condensates of pseudoscalar as well as scalar channel, we have studied another path of chiral transition. We have seen that such phase appears near the phase boundary of chiral transition, and thereby restoration of chiral symmetry is delayed to higher densities or temperatures. We can also see that magnetization in this phase shows an oscillating shape like SDW. This phase may then be characterized by local ferromagnetism and global anti-ferromagnetism.

\section{Chiral symmetry and spin density wave}

First we see the appearance of spin density wave in relation to the chiral transition. Restoration of chiral symmetry is important at moderate densities or finite temperature and many studies including lattice gauge simulations or effective model studies have been done to figure out the QCD phase diagram on the temperature-density plane \cite{wam}. Assuming non vanishing pseudosclar condensate as well as scalar one, we consider the following non-uniform configuration 
\footnote{We only consider the chiral limit in the following.}
:
\begin{figure}[h]
\begin{minipage}{0.48\textwidth}
\begin{eqnarray}
\langle\bar{\psi}\psi\rangle&=&\Delta\cos{\bf q}\cdot{\bf r}\nonumber\\
\langle\bar{\psi}i\gamma_5\tau_3\psi\rangle&=&\Delta\sin{\bf q}\cdot{\bf r}.
\end{eqnarray}
\end{minipage}
\hspace{\fill}
\begin{minipage}{0.48\textwidth}
%\begin{figure}
\begin{center}
\includegraphics[width=4cm]{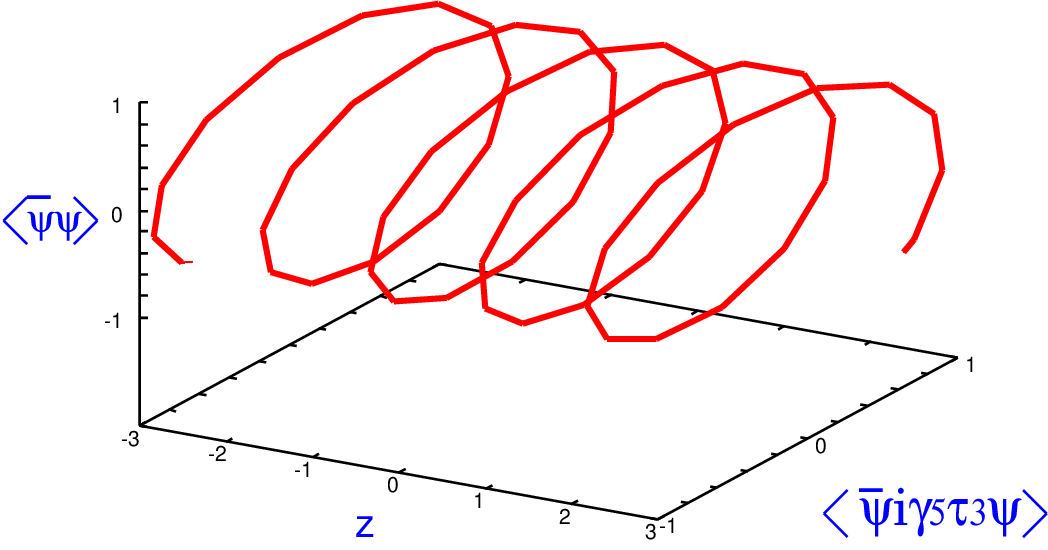}
\caption{}
\label{Fig:diagram}
\end{center}
%\end{figure}
\end{minipage}
\end{figure}
Note that this configuration respects $SU(2)\times SU(2)$ chiral symmetry and the system is a charge eigenstate. It is called dual chiral density wave (DCDW) (Fig.~1). Both condensates construct the complex order parameter, in cotrast to the usual discussion of chiral transition with uniform scalar condensate; we shall see that the amplitude $\Delta$ gives an effective mass for quarks and the phase degree of freedom $\theta={\bf q}\cdot{\bf r}$ gives rise to a magnetic property. Similar ideas of chiral density wave have been proposed as relevant phases in the large $N_c$ limit \cite{chi}, where only non-uniform scalar condensate has been considered.

Some results have been presented in Figs.~3,4 by using NJL model as an effective model of QCD at moderate densities \cite{dcdw}. Different from the usual result denoted by the thin dotted curve, DCDW appears around the critical density with finite momentum $q$ (Fig.~1). Chiral restoration is then delayed by the appearance of DCDW. The magnitude of $q$ is $O(2k_F)$, which suggests that nesting effect of the Fermi surface should be responsible to appearance of DCDW. Actually we can show that the correlation function $C(q)=\lim_{\omega\rightarrow 0}F.T.\langle\bar\psi i\gamma_5\tau_3 \psi(x)\bar\psi i\gamma_5\tau_3 \psi(0)\rangle$ diverges at finite $q$ with $O(2k_F)$ at the critical density \cite{dcdw}.

\begin{figure}[h]
\begin{minipage}{0.48\textwidth}
\begin{center}
\includegraphics[width=5cm]{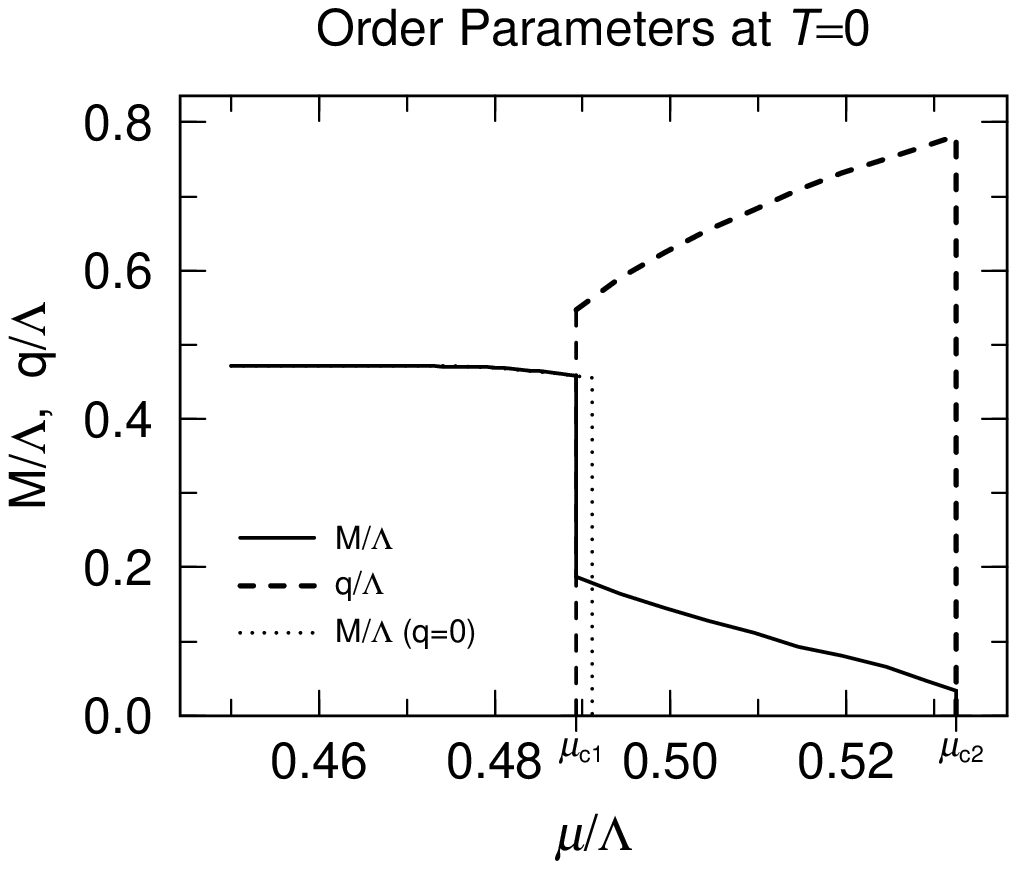}
\end{center}
\caption{Wave number $q$ and the dynamical mass $M=2G\Delta$ are plotted 
as functions of the chemical potential at $T=0$. 
Solid (dotted) line for $M$ with (without) the density wave, 
and dashed line for $q$.}
\label{op1}
\end{minipage}
\hspace{\fill}
\begin{minipage}{0.48\textwidth}
\vspace*{-0.4cm}
\begin{center}
\includegraphics[width=5cm]{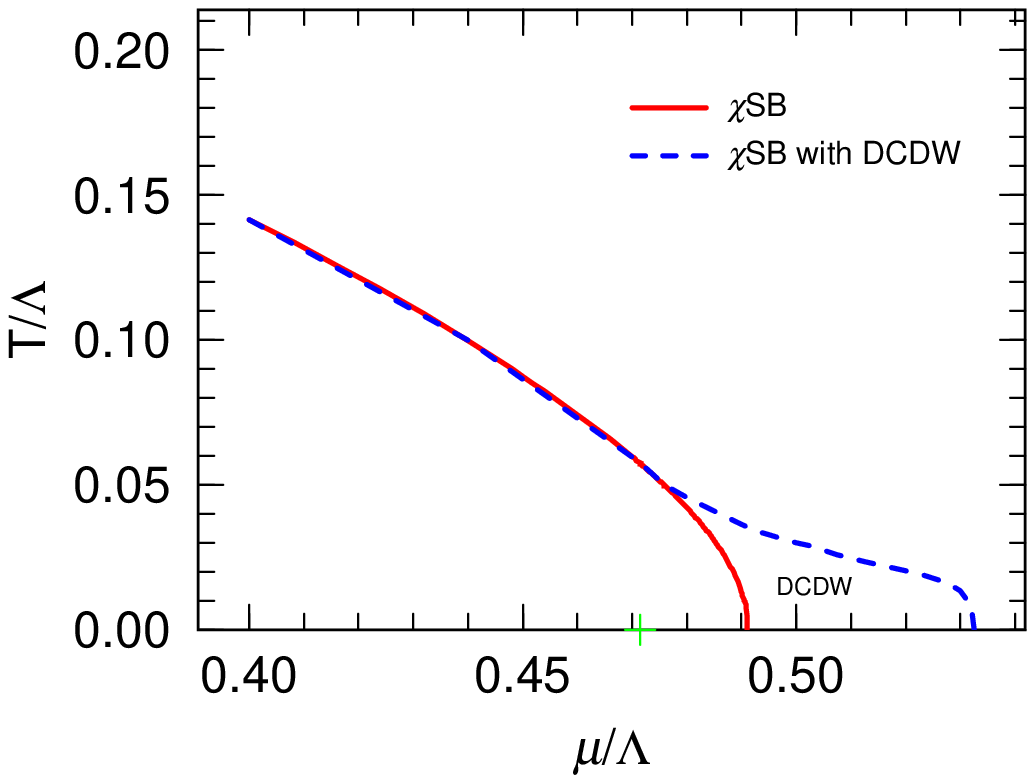}
\end{center}
%\vspace*{-0.5cm}
\caption{Phase diagram of chiral transition on the temperature-density plane. DCDW appears at relatively low temperature.}
\label{rhoCDW}
\end{minipage}
\end{figure}

Accordingly magnetization of quark matter exhibits  
\begin{equation}
\langle \bar \psi\Sigma_z \psi\rangle=M\cos({\bf q}\cdot{\bf r}),
\end{equation}
which implies there develops SDW in the DCDW phase.

It should be interesting to see a similarity to pion condensation in hadronic matter, where nucleons take a anti-ferromagnetic spin ordering in the presence of classical pion field \cite{pio}. 

\section{Ferromagnetism and magnetic susceptibility}

In the first study about ferromagnetic instability in QCD we calculated the 
energy of the spin polarized quark matter. Since quark matter is color neutral as a whole, the Fock exchange energy gives the leading-order contribution. Using the relation,  $\langle\lambda_i\rangle_{ab}\langle\lambda_i\rangle_{ba}=1/2-1/(2N_c)\delta_{ab}$, we can see that it repulsively works for any quark pair.  
%Thus the situation is similar to the electron gas interacting with the repulsive Coulomb interaction. 
Then two particles with the same spin can avoid the repulsive interaction due to the Pauli principle to favor ferromagnetic order. On the other hand the kinetic energy is totally increased. So when the energy 
gain in their interaction exceeds the increase of the kinetic energy, we can expect a ferromagnetic instability. This is the Bloch mechanism \cite{blo}. A calculation has been done by using the one-gluon-exchange interaction to find a weakly first-order phase transition around the nuclear density. 

To get more insight into the magnetic properties of quark matter we have recently studied magnetic susceptibility within Fermi-liquid theory \cite{tat082,tat083}
\footnote{We assume here the second order phase transition, but we shall find the similar critical density to the one in ref..}
. By applying tiny and uniform magnetic field $B$ we examine the magnetization $\langle M\rangle$ of quark matter to evaluate magnetic susceptibility,
$ 
\chi_M=\partial\langle M \rangle/\partial B|_{N,T,B=0}, 
$
which can be expressed in terms of the Landau-Migdal parameters:
\begin{equation}
\chi_M=\left(\frac{g_D\mu_q}{2}\right)^2/\left(\frac{\pi^2}{N_ck_FE_F}-\frac{1}{3}f_1^s+\bar{f}^a\right),
\label{chim}
\end{equation}
where $f_1^s,\bar{f}^a$ are spin-independent and -dependent Landau-Migdal parameters, respectively.

\subsection{Screening effects for gluons}

Landau-Migdal parameters usually includes infrared (IR) divergences in gauge theories QCD/QED, so that it is essential to take into account the screening effect to improve them. The HDL resummation can be achieved by using the quark polarization operator; longitudinal gluons are statically screened in terms of the Debye mass, while transverse gluons are only dynamically screened due to the Landau damping. Thus Debye screening surely improves the IR divergence for longitudinal gluons, while there still remains the IR divergences coming from transverse gluons. At $T=0$ these divergences cancel each other in (\ref{chim}) to give a meaningful result. We shall see an interesting effect caused by the dynamic screening in section 3.3.

\subsection{Magnetic transition at $T=0$}

The magnetic susceptibility is given in Fig.~4 at $T=0$.
\begin{figure}[h]
\begin{minipage}{0.6\textwidth}
\begin{center}
\includegraphics[width=6cm]{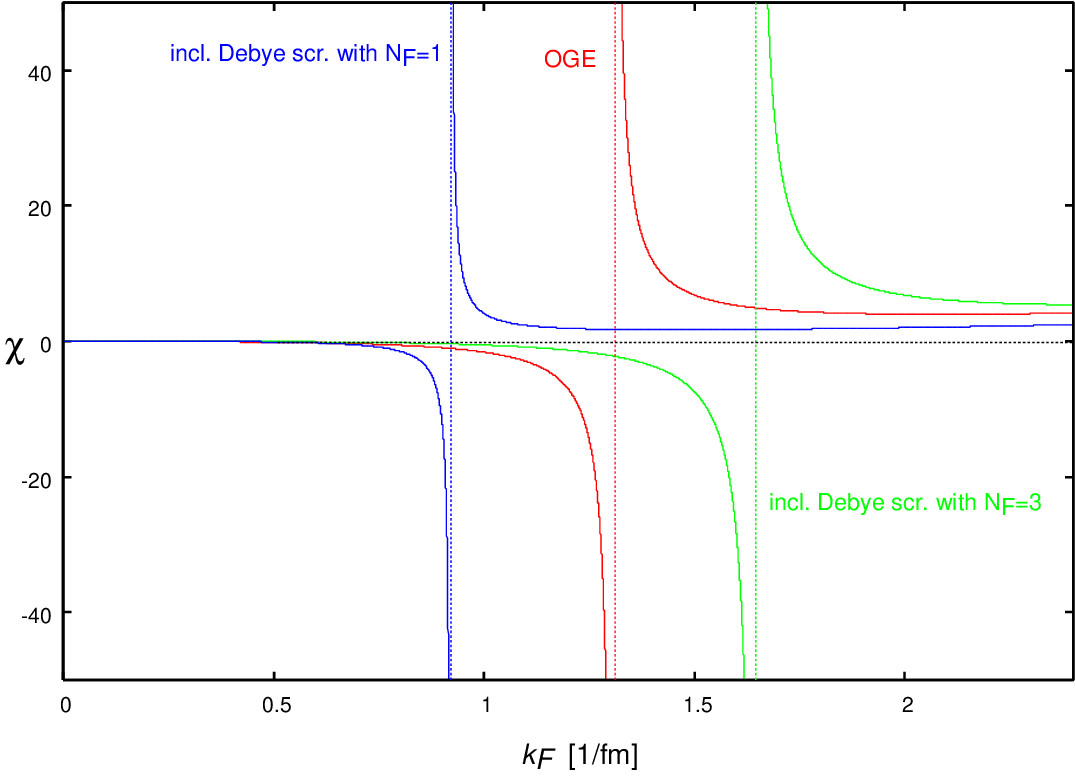}
\caption{Magnetic susceptibility at $T=0$. The solid curve shows the result
 with the simple OGE without screening, while the dashed and dash-dotted ones 
 shows the screening effect with $N_f=1$ (only $s$ quark)and $N_f=2+1$
 ($u,d,s$ quarks), respectively.}
\end{center}
\end{minipage}
\hspace{\fill}
\begin{minipage}{0.35\textwidth}
\begin{center}
\vspace{-0.8cm}
\includegraphics[width=5.cm]{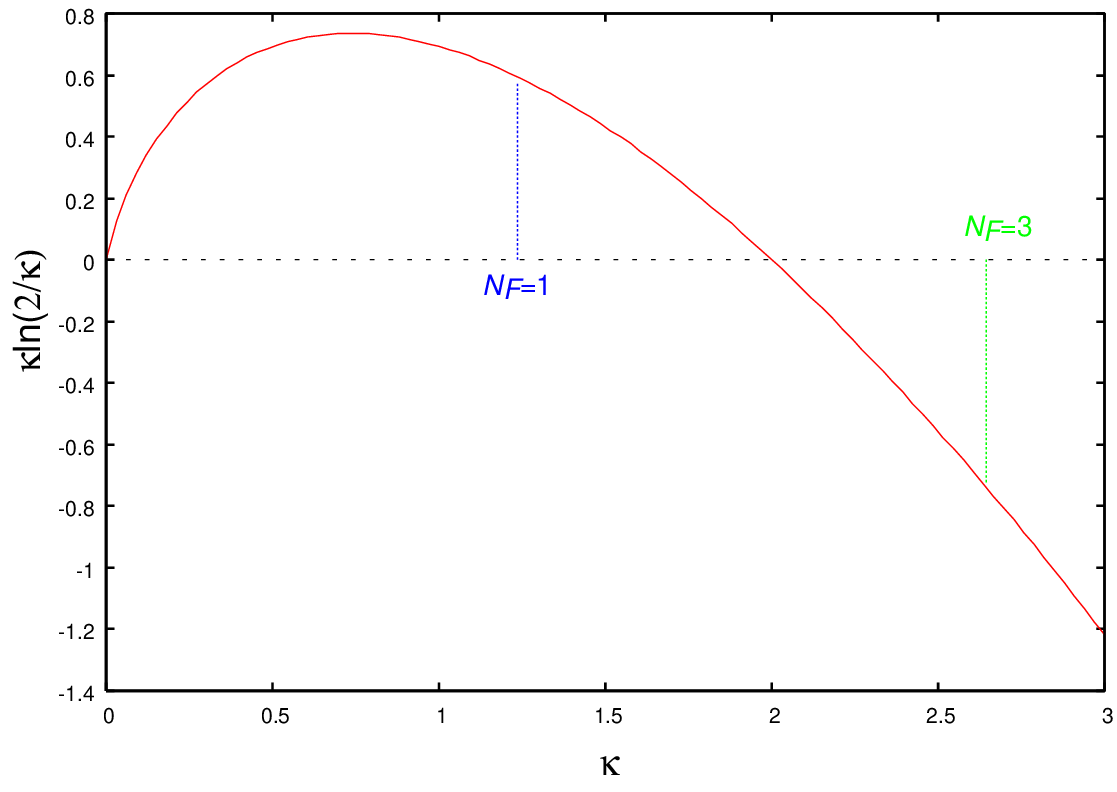}
\vspace{0.5cm}
\caption{Flavor dependence of the contribution of the screening effects.}
\end{center}
\end{minipage}
\end{figure}
We can see that magnetic susceptibility diverges around the nuclear density and  quark matter is in the ferromagnetic phase below the critical density. When the screening effects are taken into account, the curve is shifted in two ways, depending on the number of flavors; it shifted to lower densities for $N_f=1$, while to higher densities for $N_f=3$. Thus the screening effects favors the ferromagnetic phase for $N_f=3$, different from the case of $N_f=1$. This is in contrast with the usual argument for electron gas, where the correlation effect is always disfavors the magnetic transition \cite{bru57}. Such behavior can be seen by looking at the contribution of the screening effects to $\chi_M$ (Fig.~5), which reads
\begin{equation}
\Delta\chi_M^{-1}\propto \kappa\ln(2/\kappa),
\end{equation} 
with $\kappa=m_D^2/2k_F^2$, where the Debye screening mass can be written as 
$m_D^2=\sum_{\rm flavors}g^2/2\pi^2k_{F,i}E_{F,i}$. Thus the screening effect in quark matter is qualitatively different from electron gas.

\subsection{Finite temperature effects and non-Fermi-liquid behavior}

Our framework can be easily extended to finite temperature case \cite{tat083}. We, hereafter, consider the low temperature case ($T/\mu\ll 1$), but the usual low-T expansion cannot be applied since quark energy exhibits an anomalous behavior near the Fermi surface. Actually the one-loop result for the quark self-energy can be given as
\begin{equation}
{\rm Re}\Sigma_+(\omega)\simeq{\rm Re}\Sigma_+(\mu)
-\frac{C_fg^2v_F}{12\pi^2}(\omega-\mu)\ln \frac{m_D}{|\omega-\mu|}+\Delta^{\rm reg}(\omega-\mu) 
\end{equation} 
near the Fermi surface, with $v_F$ being the Fermi velocity and $C_f=(N_c^2-1)/2N_c$ \cite{man}. The second term appears due to the transverse gluons and logarithmically diverges at the Fermi surface, so that quark matter behaves as the {\it marginal} Fermi liquid \cite{smi}. 

The temperature dependent term in $\chi_M$ is finally given by 
\begin{eqnarray}
%\chi^{-1}_M(T)&=&\chi^{-1}_M(0)+\delta\chi^{-1}_M(T),\nonumber\\
\delta\chi_M^{-1}&=&\chi_{\rm Pauli}^{-1}\Big[ \frac{\pi^2}{6k_F^4}\left(2E_F^2-m^2+\frac{m^4}{E_F^2}\right)T^2 \notag\\
&&+\frac{C_fg^2v_F}{72k_F^4E_F^2} \left(2k_F^4+k_F^2m^2+m^4\right)T^2\ln\left(\frac{m_D}{T}\right)\Big]+O(g^2T^2).
\label{delT}
\end{eqnarray}
We can see that there appears $T^2\ln T$ term beside the usual $T^2$ one. This is a novel non-Fermi-liquid effect \cite{tat083}. It should be interesting to compare this result with other typical non-Fermi-liquid effect in the specific heat \cite{hol} or the gap function in color superconductivity \cite{son}. Moreover, it is to be noted that the spin fluctuation effect gives $T^3\ln T$ term as a leading-order contribution in electron gas \cite{bea}. 

Finally the magnetic phase digram is presented on the temperature-density plane (Fig.~5), where we can see the Curie temperature of several tens MeV.

\begin{figure}[h]
\begin{center}
\includegraphics[width=0.45\textwidth]{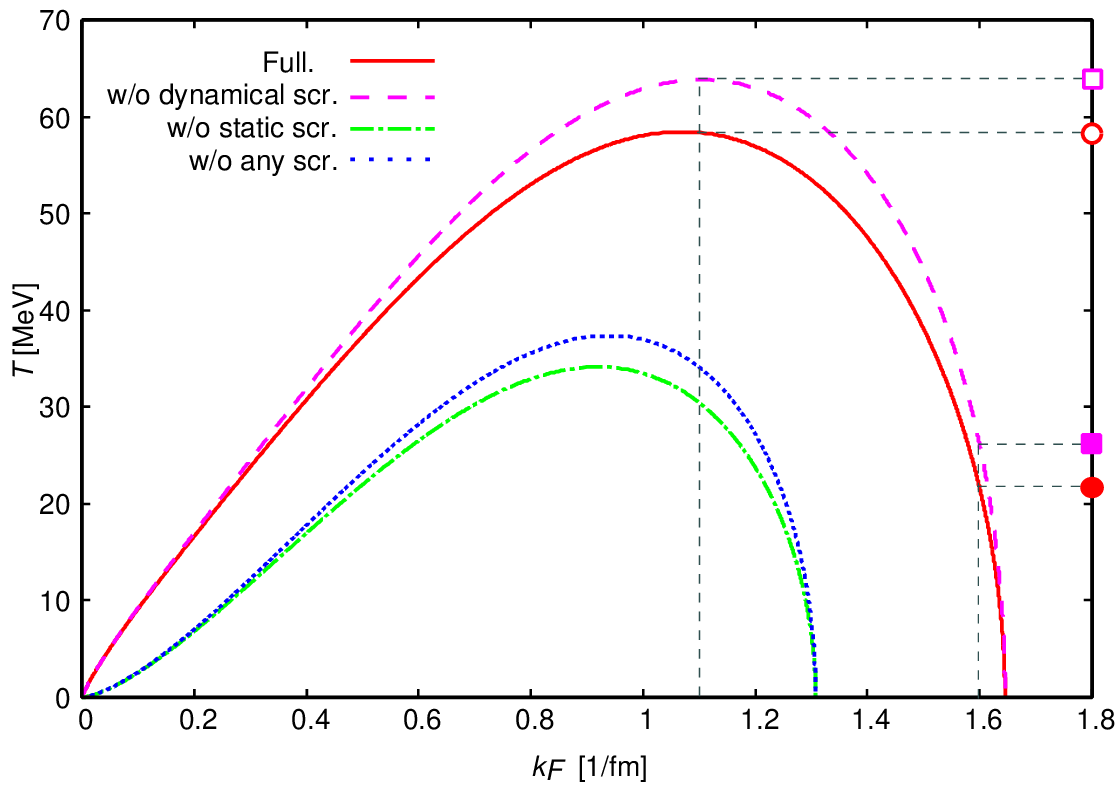}
\caption{Magnetic phase diagram in the density-temperature plane. The
 solid, dashed, dash-dotted, dotted curves show the results for the full
 expression, the one without the $T^2 \ln T$ term,
 without the $\kappa \ln \kappa$ term, and without the $T^2 \ln T$ and
 $\kappa \ln \kappa$ terms. The open (filled)
 circle indicates 
the Curie temperature at $k_F=1.1(1.6)$ fm$^{-1}$ while the squares show those when we disregard the $T^2 \ln T$ dependence.}
\label{Fig:diagram}
\end{center}
\end{figure}

\section{Summary and concluding remarks}

We have discussed some magnetic aspects of QCD on the temperature-density plane.First we have demonstrated appearance of DCDW near the phase boundary of chirak transition.
In recent papers Nickel discussed the appearance of the real kink crystal (RKC) \cite{nic}. The tricritical point is then Lifshitz point in this case. This is an interesting possibility, but more studies are needed to elucidate the relation between RKC and DCDW phases, while he cocluded that RKC is more favored than DCDW phase. 

We have studied the static magnetic susceptibility of quark matter by utilizing  Fermi-liquid theory to see ferromagnetic transition, where the screening effects for gluon propagators become very important; the static screening by the Debye mass gives $g^4\ln g^2$ term at $T=0$, while it produces a novel non-Fermi liquid effect by the $T^2\ln T$ term. For a more realistic study, however, some non-perturbative effects are taken into account at moderate densities. Moreover, the restoration of chiral symmetry is also important there. 

Observational signals of the magnetic phases may be found out in thermal evolutions of compact stars, besides the direct evidence of magnetic evolution. Nambu-Goldstone bosons, as a consequence of spontaneous symmetry breaking in the magnetic phases, may contribute to thermodynamical quantities to modify their thermal evolutions. 

\section*{Acknowledgements}

%We would like to thank ...........
This work was partially supported by the 
Grant-in-Aid for the Global COE Program 
``The Next Generation of Physics, Spun from Universality and Emergence''
from the Ministry of Education, Culture, Sports, Science and Technology
(MEXT) of Japan  
and the 
Grant-in-Aid for Scientific Research (C) (16540246, 20540267).

%\appendix
%\section{First Appendix} %Empty argument \section{} yields `Appendix'. 
%
%\section{Second Appendix}

\end{document}